\documentclass[11pt]{article}
\usepackage{osid}
\usepackage{amsmath}
\usepackage{amssymb}
\usepackage{latexsym}
\usepackage{cmmib57}
\usepackage[utf8]{inputenc}
\usepackage{graphicx}

\def\idty{{\leavevmode{\rm 1\ifmmode\mkern -4.8mu\else\kern -.3em\fi
      I}}}

\title{Distilling entanglement from Fermions}
\author{Michael Keyl\\
  {\footnotesize ISI Foundation, Viale S. Severo 65, 10133 Torino, Italy,
    e-mail: keyl@isi.it}}

\def\RHD{M. Keyl, Distilling entanglement from Fermions}

\begin{document}
\maketitle

\begin{abstract}
  Since Fermions are based on anti-commutation relations, their entanglement can
  not be studied in the usual way, such that the available theory has to be
  modified appropriately. Recent publications consider in particular the
  structure of separable and of maximally entangled states. In this talk we want
  to discuss local operations and entanglement distillation from bipartite,
  Fermionic systems. To this end we apply an algebraic point of view where
  algebras of local observables, rather than tensor product Hilbert spaces play
  the central role. We apply our scheme in particular to Fermionic Gaussian states
  where the whole discussion can be reduced to properties of the covariance
  matrix. Finally the results are demonstrated with free Fermions on an infinite,
  one-dimensional lattice.
\end{abstract}

\section{Introduction}

Entanglement distillation is one of the most fundamental processes of quantum
information processing \cite{DPGBook}. It is an integral part of many
protocols and devices like quantum repeaters  and it provides important
procedures to measure the entanglement content of a given physical system. In
the usual setup it is assumed that two distant parties -- Alice and Bob --
share a large number of (weakly) entangled pairs of particles, and the task is
to generate a (possibly small) amount of highly (or even maximally) entangled
pairs by means of local operations and classical communication. In this
context it is implicitly assumed that the particles shared by Alice and Bob
are \emph{distinguishable}: Firstly there is a clear distinction between
particles controlled by Alice and those controlled by Bob. Secondly many
distillation protocols require the selection of particular pairs by Alice and
Bob (e.g. to perform filtering operations). Hence even the local particles
needs to be distinguishable. 

At a first glance it therefore appears to be completely impossible to talk
about entanglement distillation with fermions, because the latter are by
definition \emph{undistinguishable}. A second thought reveals, however, very
quickly that this is only an apparent difficulty. The only thing we have to
drop is the focus on particles. Instead we have to consider setups where Alice
and Bob control independent subsystems (usually distinguished by their position
in space) of a larger physical system. A typical example is a Fermi gas from
which Alice and Bob want to distill entanglement by using only operations
(together with classical communication) which are localized in spatially
separated regions (e.g. Alice's and Bob's laboratories).

Mathematically such a situation is most easily described in terms of operator
algebras. In other words, instead of using tensor products of Hilbert spaces,
we describe bipartite systems by specifying which local observables are
measurable by Alice and which by Bob. This approach is successfully applied
to the study of entanglement of infinite degrees of freedom systems
\cite{MR2153773,KMSW06} and for the analysis of separable \cite{Mo06,BaCiWo07}
and maximally entangled \cite{MEF08} states of Fermionic systems (cf. also the
references in \cite{BaCiWo07} for more literature on Fermionic entanglement). 

The purpose of this paper is to study entanglement distillation in the same
framework. In this context we will show that distillation from Fermions can be
treated basically in the same way as ordinary distillation with only some
small changes which mainly arise from the emergence of super selection
rules. In addition we will present an explicit scheme which can be applied to
any quasi-free state and which allows explicit calculations (e.g. of
distillation rates) for fairly large systems, such as (subsystems of) infinite
quasi-free lattice models. 

\section{Entanglement distillation}
\label{sec:entangl-dist}
 
Let us start with a short look on standard distillation techniques. Hence
assume that Alice and Bob share $N$ $d$-level systems in the joint state
$\rho^{\otimes N}$, where $\rho$ denotes a density matrix on the Hilbert space
$\mathcal{H} \otimes \mathcal{H}$, $\mathcal{H}=\mathbb{C}^d$. To generate maximally
entangled qubit pairs from these resources they can proceed as follows: 
\begin{enumerate}
\item 
  Look for (and drop) unentangled subsystems. Mathematically this means to find
  a unitary $U: \mathcal{H} \rightarrow \mathcal{H}_1 \otimes \mathcal{K}$,
  $\mathcal{H}_1 = \mathbb{C}^{d_1}$, and to apply the transformation
  $T_1(\rho) = \operatorname{tr}_{\mathcal{K}}(U \rho U^*)$. The final state
  $\rho_1$ should contain (almost) as much entanglement as the original
  $\rho$. In other words: $\mathcal{K}$ and $U$ has to be chosen appropriately.   
\item  
  Find a maximally entangled state $\psi \in \mathcal{H}_1 \otimes
  \mathcal{H}_1$ such that 
  \begin{equation} \label{eq:1}
     \langle \psi, \rho_1 \psi \rangle > d_1^{-1}.
  \end{equation}
 The best choice would be to take the $\psi$ which maximizes this fidelity. To
 be successful here the first step  is in many cases mandatory, because a
 state can be entangled without satisfying inequality (\ref{eq:1})  for any
 maximally entangled $\psi$. This can  be easily seen if we choose
 $d=2\tilde{d}$ and 
 \begin{equation}
   \rho = |\phi\rangle\langle\phi| \otimes \frac{\idty}{\tilde{d}^2} \quad
   \phi = 2^{-1/2} (|00\rangle+|11\rangle) 
 \end{equation}
 since we get $\sup_\psi \langle\psi,\rho \psi\rangle = 1/\tilde{d}^2$, which
 can be arbitrarily small (if $\tilde{d}$ is big enough) although $\rho$ is
 distillable.  
\item 
  Consider the group
  \begin{equation}
    G_\psi = \{ U_A \otimes U_B\, | \,
    U_A, U_B \in \operatorname{U}(d_1),\ U_A \otimes U_B \psi = \psi \}, 
  \end{equation}
  and average over it. This leads to the \emph{twirl} operation given by
  \begin{equation}
    T_2(\rho_1) = \int_{G_\psi} U \rho U^* dU,
  \end{equation}
  where $dU$ denotes the Haar measure on $G_\psi$. 
\item 
  The output of the channel $T_2$ is an isotropic state 
  \begin{equation}
    \rho_2 = T_2 \rho_1 = \vartheta |\psi\rangle\langle\psi| + 
    (1-\vartheta) \frac{\idty}{d_1^2} 
  \end{equation}
  with $\vartheta \in [- (d_1^2-1)^{-1}, 1]$ given in terms of the fidelity
  $f = \langle \psi, \rho \psi\rangle$ by   
  \begin{equation}
    \vartheta = \frac{d_1^2 f-1}{d_1^2-1}.
  \end{equation}
  From Equation (\ref{eq:1}) it follows immediately that $\rho_2$ again
  satisfies the inequality 
  \begin{equation}
    \langle\psi, \rho_2 \psi \rangle > d_1^{-1},
  \end{equation}
  and therefore it is distillable \cite{H2RedKrit}.
\item 
  Now we can continue with standard techniques for isotropic states which
  provide us with a number $M$ of (almost) maximally entangled \emph{qubit}
  pairs; cf \cite{VoWo03} and the references therein. 
\end{enumerate}

\section{Bipartite Fermionic systems}
\label{sec:fermionic-systems}

To study entanglement of Fermionic systems the usual framework which relies on
tensor product Hilbert spaces is too narrow because indistinguishability and
anti-commutation relations has to be taken into account. Instead, it is more
appropriate to describe the splitting of the overall system into two
subsystems in terms of observables algebras. This approach was successfully
applied in particular to infinite degrees of freedom systems
\cite{MR2153773,KMSW06}. For our purposes a simplified (finite dimensional)
approach is sufficient.

\begin{definition}{Definition} \label{def:1}
  A bipartite quantum system consists of a Hilbert space and two C*-algebras
  $\mathcal{A}, \mathcal{B} \subset \mathcal{B}(\mathcal{H})$ which commute
  elementwise (i.e. $[A,B]=0$ $\forall A \in \mathcal{A}$, $\forall B \in
  \mathcal{B}$).   
\end{definition}

Selfadjoint elements of $\mathcal{A}$ and $\mathcal{B}$ describe the
(projection valued) observables of the given system which can be
\emph{locally} measured by Alice and Bob respectively. The default setup can
be recovered if we have  
\begin{equation} \label{eq:3}
  \mathcal{A} = \mathcal{B}(\mathcal{H}_A) \otimes \idty_B,\quad
  \mathcal{B}=\idty_A \otimes \mathcal{B}(\mathcal{H}_B), \quad  
  \mathcal{M} = \mathcal{B}(\mathcal{H}_A \otimes \mathcal{H}_B)
\end{equation}
in terms of two Hilbert spaces $\mathcal{H}_A, \mathcal{H}_B$. Note, however,
that we have neither assumed that $\mathcal{A}$ and $\mathcal{B}$ together
generate $\mathcal{B}(\mathcal{H})$ nor that $\mathcal{B}$ is the commutant of
$\mathcal{A}$ (in contrast to \cite{KMSW06}). Therefore beside (\ref{eq:3})
other realizations of bipartite systems are possible, even if $\mathcal{H}$ is
finite dimensional. A particular example arises, if $\mathcal{H}_{A/B}$
decomposes into a direct sum $\mathcal{H}_{A/B} = \mathcal{H}_{A/B}^+ \oplus
\mathcal{H}_{A/B}^-$, and if we define 
\begin{equation} \label{eq:27}
  \mathcal{A} = \bigl(\mathcal{B}(\mathcal{H}_A^+) \oplus
  \mathcal{B}(\mathcal{H}_A^-) \bigr) \otimes \idty_B, \quad 
  \mathcal{B} = \idty_B \otimes \bigl(\mathcal{B}(\mathcal{H}_A^+) \oplus 
  \mathcal{B}(\mathcal{H}_A^-) \bigr). 
\end{equation}
This is -- as we will see -- exactly the situation we have to study for a
system consisting of a finite number of Fermions. 
 
To explain the latter remark consider the Hilbert space $\mathcal{K}$ and the
corresponding antisymmetric Fockspace $\mathcal{H} =
\mathcal{F}_-(\mathcal{K})$. For each $h, f \in  \mathcal{K}$ we can define
the usual creation and annihilation operators $c^*(h)$ and $c(f)$ on
$\mathcal{H}$, which satisfy the canonical anti-commutation relations
($\{\,\cdot\,,\,\cdot\,\}$ denotes the anti-commutator)  
\begin{equation} \label{eq:5}
  \{ c(f), c(h) \} = \{ c^*(f), c^*(h)\} = 0,\quad \{c^*(h), c(f)\} = \langle 
  h,f \rangle \idty.
\end{equation}
In some cases it is more convenient to combine $c$ and $c^*$ in one
operator (this is called the self-dual formalism \cite{MR0295702,Araki87}):
\begin{equation}\label{eq:28}
  B(f,h) = c(f) + c^*(\overline{h}),\quad \Gamma (f,h) =
  (\overline{h},\overline{f}),  
\end{equation}
where $(\overline{\,\cdot\,})$ denotes complex conjugation in an appropriately
chosen (and fixed!) basis. Now we get from (\ref{eq:5})
\begin{equation}
  \{ B(F_1), B(F_2) \} = \langle \Gamma F_1, F_2\rangle,\quad B(F)^* =
  B(\Gamma F), \forall F, F_1, F_2 \in \mathcal{K} \oplus \mathcal{K}. 
\end{equation}
The set of operators $\{c(f) \, | \, f \in \mathcal{K} \} \subset
\mathcal{B}(\mathcal{H})$ generates a C*-algebra
$\operatorname{CAR}(\mathcal{K}) \subset \mathcal{B}(\mathcal{H})$ which is
called the algebra of canonical ani-commutation relations. It can be regarded
as the closure (in operator norm) of the algebra of polynomials in the $c(f)$
and $c^*(h)$.  

Now consider the parity operator on $\mathcal{H}$ which is given in terms of the
number operator $N$ by $\theta = (-1)^N$. It acts as the identity on the
subspace $\mathcal{H}^+ \subset \mathcal{H}$ of vectors with an even particle number
and as minus the identity on the complementary subspace $\mathcal{H}^-$. We write
$P^\pm$ for the corresponding projections and get 
\begin{equation}
  P^\pm : \mathcal{H} \rightarrow \mathcal{H}^\pm,\quad \theta = P^+ - P^-.
\end{equation}
 The elements of $\operatorname{CAR}(\mathcal{K})$ which commute with $\theta$
 are called \emph{even elements} and they  form the \emph{even subalgebra}  
\begin{equation}
  \operatorname{CAR}_+(\mathcal{K}) = 
  \{ A \in \operatorname{CAR}(\mathcal{K}) \, | \, [A, \theta] = 0 \}  
\end{equation}
of $\operatorname{CAR}(\mathcal{K})$. It can be regarded as the (closure of)
the algebra of \emph{even} polynomials in $c(f)$ and $c^*(h)$. 

If $\mathcal{K}$ is finite dimensional $\operatorname{CAR}(\mathcal{K})$
coincides with $\mathcal{B}(\mathcal{H})$, i.e. it is a full matrix
algebra. The even subalgebra however is given by  
\begin{equation} \label{eq:8}
  \operatorname{CAR}_+(\mathcal{K}) = 
  \mathcal{B}(\mathcal{H}^+) \oplus \mathcal{B}(\mathcal{H}^-). 
\end{equation}
This can be easily seen from the fact that a product of an \emph{even number}
of creation and annihilation operators can change the particle number only by
a factor of 2.  

The next step is to decompose $\mathcal{K}$ into an ``Alice'' and a ``Bob''
subspace, i.e. $\mathcal{K} = \mathcal{K}_A \oplus \mathcal{K}_B$. Then we can
associate to $\mathcal{K}_{A/B}$ the corresponding Fockspaces
$\mathcal{H}_{A/B} = \mathcal{F}_-(\mathcal{K}_{A/B})$, and also the
CAR-algebras $\operatorname{CAR}(\mathcal{K}_{A/B})$ and
$\operatorname{CAR}_+(\mathcal{K}_{A/B})$. Obviously we have   
\begin{equation}
  \mathcal{H} = \mathcal{F}_-(\mathcal{K}) \cong \mathcal{F}_-(\mathcal{K}_A)
  \otimes \mathcal{F}_-(\mathcal{K}_B) = \mathcal{H}_A \otimes \mathcal{H}_B,
\end{equation}
and similarly $\operatorname{CAR}(\mathcal{K})$ is isomorphic to
$\operatorname{CAR}(\mathcal{K}_A) \otimes \operatorname{CAR}(\mathcal{K}_B)$
if we consider the spatial tensor product. The corresponding isomorphism
satisfies   
\begin{equation} \label{eq:6}
  c(f_A \oplus 0) \mapsto c_A(f_A) \otimes \idty_B,\quad 
  c(0 \oplus f_B) \mapsto \theta_B \otimes c_B(F_B),
\end{equation}
where the $c_{A/B}(f_{A/B})$ denote the annihilation operators and
$\theta_{A/B}$ the parity operators on $\mathcal{H}_{A/B}$.  The latter are
related to the global parity $\theta$ by  
\begin{equation} \label{eq:7}
  \theta = \theta_A \otimes \theta_B.
\end{equation}
Equation (\ref{eq:6}) shows that $\operatorname{CAR}(\mathcal{K}_{A/B})$ can
not both be embedded as tensor factors into $\operatorname{CAR}(\mathcal{K})$
without violating the anti-commutation relations. The construction given in
(\ref{eq:6}) is therefore called the \emph{twisted} tensor product of
$\operatorname{CAR}(\mathcal{K}_A)$ and $\operatorname{CAR}(\mathcal{K}_B)$.  

The discussion of the last paragraph has shown that two operators $A \in
\operatorname{CAR}(\mathcal{K}_A)$, $B \in \operatorname{CAR}(\mathcal{K}_B)$
do not commute in general and therefore these two algebras can not be chosen
as the observable algebras $\mathcal{A}$ and $\mathcal{B}$. If we choose,
however, $A$ and $B$ to be even elements we immediately get from (\ref{eq:6})
that $[A,B]=0$ holds. Hence 
\begin{equation}
  \mathcal{A} = \operatorname{CAR}_+(\mathcal{K}_A),\quad
  \mathcal{B}=\operatorname{CAR}_+(\mathcal{K}_B) 
\end{equation}
is an appropriate choice for the \emph{local} observable algebras of Alice and
Bob. Together they define a bipartite Fermionic system in the sense of
Definition \ref{def:1} If $\mathcal{K}_A$ and $\mathcal{K}_B$ are finite dimensional
we can insert Equation (\ref{eq:8}) for $\mathcal{A}$ and $\mathcal{B}$ and we recover
the example already given in (\ref{eq:3}). Note in addition that $\mathcal{A}
\subset \mathcal{B}(\mathcal{H}_A)$ and $\mathcal{B} \subset
\mathcal{B}(\mathcal{H}_B)$ holds, while the \emph{full} CAR algebras fail to
have this property -- by virtue of Equation (\ref{eq:6}). Hence it is
reasonable to consider $\mathcal{H}_A$ ($\mathcal{H}_B$) as Alice's (Bob's)
Hilbert space.  

\section{Local operations}
\label{sec:local-operations}

On top of the scheme described in the last section all basic notions of
entanglement theory can be reconstructed. This was done for separable states
in \cite{Mo06,BaCiWo07} and for maximally entangled states in \cite{MEF08}. To
discuss entanglement distillation we need the concept of a local operation
which can be defined as follows (cf. \cite{KMSW06}):

\begin{definition}{Definition} \label{def:2}
  Consider two bipartite systems $\mathcal{A}_j, \mathcal{B}_j \subset
  \mathcal{B}(\mathcal{H}_j)$, $j=1,2$. An operation (i.e. a completely
  positive map) $T: \mathcal{B}(\mathcal{H}_1) \rightarrow
  \mathcal{B}(\mathcal{H}_2)$ is called 
  \emph{local} if  
  \begin{equation}
    T(\mathcal{A}_1) \subset \mathcal{A}_2,\quad 
    T(\mathcal{B}_1) \subset \mathcal{B}_2 
  \end{equation}
  and
  \begin{equation} \label{eq:4}
    T(AB) = T(A)T(B) \quad 
    \forall A \in \mathcal{A}_1,\ \forall B \in \mathcal{B}_1 
  \end{equation}
  holds.
\end{definition}

In the standard framework (\ref{eq:3}) with finite dimensional Hilbert spaces
$\mathcal{H}_A, \mathcal{H}_B$ (or if we assume that the operation $T$ is
normal) this definition coincides with the usual one. Note that the
factorization condition (\ref{eq:4}) is needed to make this statement true
\cite{KMSW06}.  

To generalize the distillation protocol from Section \ref{sec:entangl-dist}
only a few special local operations are needed. They are summarized in the
following list.

\begin{itemize}
\item 
  \emph{Local unitaries.} The easiest case is a local unitary transformation
  $A \mapsto U^* A U$ with $U = U_A \otimes U_B$ and $U_A$ ($U_B$) unitary on
  $\mathcal{H}_A$ ($\mathcal{H}_B$). It is easy to see that $U_A^* \mathcal{A}
  U_A = \mathcal{A}$ is equivalent to $U_A \mathcal{H}_A^{\pm} =
  \mathcal{H}_A^{\pm}$ or $U_A \mathcal{H}_A^{\pm} = \mathcal{H}_A^{\mp}$. A
  similar statement holds for $U_B$. 
\item 
  \emph{Local Bogolubov transformations.} A special case of the previous
  example arises if $U_A$ and $U_B$ are related to unitaries $u_A$, $u_B$ on
  $\mathcal{K}_A \oplus \mathcal{K}_A$, $\mathcal{K}_B \oplus \mathcal{K}_B$ by 
  \begin{equation} \label{eq:29}
    U^* B(F) U = B(uF),\ \text{with}\ U = U_A \otimes U_B,\ u = u_A \oplus u_B,
  \end{equation}
  where $B(F)$ is the operator introduced in Equation (\ref{eq:28}). It is
  easy to see that (\ref{eq:29}) is only possible if 
  \begin{equation} \label{eq:2}
     \Gamma u \Gamma = u
  \end{equation}
  holds (cf. again (\ref{eq:28})). The condition (\ref{eq:2}) is on the other
  hand sufficient for the existence of a unitary $U$ satisfying (\ref{eq:29})
  for a given $u$.
\item 
  \emph{Discarding subsystems.} Consider now decompositions
  $\mathcal{K}_{A/B} = \mathcal{K}_{A/B,1} \oplus \mathcal{K}_{A/B,2}$ of
  $\mathcal{K}_A$ and $\mathcal{K}_B$. If we denote the Fockspaces of
  $\mathcal{K}_j = \mathcal{K}_{A,j} \oplus \mathcal{K}_{B,j}$, $j=1,2$ by
  $\mathcal{H}_j$ we get a decomposition of $\mathcal{H}$ into a tensor
  product $\mathcal{H} = \mathcal{H}_1 \otimes \mathcal{H}_2$. If we perform
  the partial trace over say $\mathcal{H}_2$ we get the Fockspace
  $\mathcal{H}_1$ and the corresponding reduced observable algebras
  $\mathcal{A}_1, \mathcal{B}_1 \subset \mathcal{B}(\mathcal{H}_1)$. In this
  way the partial trace becomes (the Schr\"odinger picture version of) a local
  operation between two bipartite Fermionic systems, which discards the modes
  belonging to $\mathcal{K}_2$.
\item 
  A \emph{joint parity measurement} is described by the PVM
  \begin{equation}
    P^{jk} = P_A^j \otimes P_B^k,\quad j,k = +,-
  \end{equation}
  and the corresponding von Neumann-L\"uders instrument ($P_{A/B}^{\pm}$
  denote the projections to the even/odd subspaces of $\mathcal{H}_A$ and
  $\mathcal{H}_B$; cf. Section \ref{sec:fermionic-systems}). For a system in
  the state $\rho$ the probability to get the outcome $j,k$ is $p^{jk}$ and
  the corresponding output state is $\rho^{jk}$:
  \begin{equation}
    p^{jk} = \operatorname{tr}(P^{jk} \rho),\quad  
    \rho^{jk} = \frac{P^{jk} \rho P^{jk}}{p^{jk}}. 
  \end{equation}
  The projections $P^{jk}$ commute with all $A \in \mathcal{A}$ and all $B \in
  \mathcal{B}$. Therefore parity measurements can be done without disturbing
  the system. This implies immediately that the state $\rho$ can not be
  distinguished from the mixture $\sum_{jk} p^{jk} \rho^{jk}$. Within a
  distillation scheme this instrument can be used to perform local filtering
  operations; e.g. Alice and Bob can decide to drop the whole system if their
  local parities are different and to keep it otherwise. If we set
  \begin{equation} 
    \label{eq:30} \mathcal{K}_A = \mathcal{K}_B = \mathbb{C}^d
    \Rightarrow \mathcal{H}_A^+ \cong \mathcal{H}_A^- \cong \mathcal{H}_B^+
    \cong \mathcal{H}_B^- \cong \mathbb{C}^D,\quad D=2^{d-1},
  \end{equation}
  and ignore the value of the parities (apart from $j=k$) we get a
  (non-unital) local operation which transforms a bipartite Fermionic system
  into a pair of $D-$level systems in the state
  \begin{equation}
    \frac{p^{++} \rho^{++} + p^{--} \rho^{--}}{p^{++} + p^{--}}.
  \end{equation}
\end{itemize}

\section{Distilling from Fermions}
\label{sec:dist-from-ferm}

Let us now adopt the general distillation scheme sketched in Section
\ref{sec:entangl-dist} to the Fermionic case. To this end we will use
throughout this section the assumptions made in Equation (\ref{eq:30}), which
implies in particular that (\ref{eq:27}) holds.  In addition, consider two
maximally entangled vectors $\varphi_+ \in \mathcal{H}^{++}$, $\varphi_- \in
\mathcal{H}^{--}$ and
\begin{equation} \label{eq:16} 
  \psi_{\pm} = \frac{1}{\sqrt{2}} (\varphi_+ \pm \varphi_-).
\end{equation}
For each $A \in \mathcal{A} \otimes \mathcal{B}$ we have
\begin{equation}
  \operatorname{tr}(A |\psi_{\pm}\rangle\langle\psi_{\pm}|) = 
  \frac{1}{2} \left(\operatorname{tr}(A |\varphi_+\rangle\langle\varphi_+|) + 
    \operatorname{tr}(A |\varphi_-\rangle\langle\varphi_-|)\right).  
\end{equation}
Using the terminology from Equations (\ref{eq:9}) and (\ref{eq:10}) this can
be rewritten as:
\begin{equation}
  p^{++}=p^{--}=\frac{1}{2},\ p^{+-}=p^{-+}=0,\
  \rho^{++}=|\varphi_+\rangle\langle\varphi_+|,\
  \rho^{--}=|\varphi_-\rangle\langle\varphi_-|.    
\end{equation}
Hence Alice and Bob can not distinguish the vector states
$|\psi_{\pm}\rangle\langle\psi_{\pm}|$ from themselves and from the mixture of
$|\varphi_+\rangle\langle\varphi_+|$ with
$|\varphi_-\rangle\langle\varphi_-|$. The latter is according to \cite{MEF08}
a Fermionic maximally entangled state (implying in particular that EOF is
maximal).

The only step from the list in Section \ref{sec:entangl-dist} we have to
change is the twirling, because averaging over the group $G_{\psi_+}$ (or
$G_{\psi_-}$) breaks the superselection rule and is therefore not an allowed
local operation. Instead, we have to look at the subgroup
\begin{equation}
  H_{\psi_+} = \{ U_A \otimes U_B \in G_{\psi_+}\, | \, U_A \mathcal{A} U_A =
  \mathcal{A},\quad U_B \mathcal{B} U_B = \mathcal{B}\}. 
\end{equation}
The structure of this group is given by the following Proposition

\begin{theorem}{Proposition} \label{prop:1} The group $H_{\psi_+}$ is
  generated by the subgroup
  \begin{equation}
    H_{\psi_+,0} = \{ U_A \otimes U_B \in G_{\psi_+} \, | \,  [U_A,
    \theta_A]=0,\quad [U_B, \theta_B]=0 \} 
  \end{equation}
  and $V = V_A \otimes V_B$ given by
  \begin{equation}
    V_A e_{A,j}^+ = e_{A,j}^-,\quad V_B e_{B,j}^+=e_{B,j}^- 
  \end{equation}
  where $e_{A/B,j}^{\pm}$, $j=1,\dots,D$, $D=2^{d-1}$ are given in terms of
  the Schmidt decomposition of $\varphi_{\pm}$, i.e.
  \begin{equation}
    \varphi_+ = \frac{1}{\sqrt{D}} \sum_{j=1}^D e_{A,j}^+ \otimes
    e_{B,j}^+,\quad 
    \varphi_- = \frac{1}{\sqrt{D}} \sum_{j=1}^D e_{A,j}^- \otimes e_{B,j}^- .  
  \end{equation}
\end{theorem}

\par\noindent\textit{Proof.\ }
  Obviously $H_{\psi_+,0} \subset H_{\psi_+}$ and $V_A \otimes V_B \in
  H_{\psi_+}$. To show the other inclusion recall from the discussion of local
  unitaries in the last section that $U_A \mathcal{A} U_B^* = \mathcal{A}$ is
  equivalent to $U_A \mathcal{H}_A^{\pm} = \mathcal{H}_A^{\pm}$ (i.e. $[U_A,
  \theta_A] =0$) or $U_A \mathcal{H}_A^{\pm} = \mathcal{H}_A^{\mp}$, and that
  a similar statement holds for $U_B$. The assumption $U_A \otimes U_B \psi_+
  = \psi_+$ implies in addition that $[U_A, \theta_A]=0 \Leftrightarrow
  [U_B,\theta_B] = 0$ holds. Hence $U \in H_{\psi_+}$ is either in
  $H_{\psi_+,0}$ or it can be written as $U = \tilde{U} V$ with a $\tilde{U}
  \in H_{\psi_+,0}$, which concludes the proof.
\hfill $\Box$ \medskip

Averaging over the group $H_{\psi_+}$ leads to states which are $H_{\psi_+}$
invariant. Their structure is given by the following proposition.

\begin{theorem}{Proposition}
  Each $H_{\psi_+}$-invariant state $\sigma$ can be written as
  \begin{equation} \label{eq:13} \sigma = 
    \lambda_+ |\psi_+\rangle\langle\psi_+| + 
    \lambda_- |\psi_-\rangle\langle\psi_-| +  
    \mu_+ (P^{++} + P^{--}) + \mu_-(P^{+-} + P^{-+})
  \end{equation}
  with
  \begin{gather} \label{eq:14} p^{++} = p^{--} = \frac{\lambda_+ +
      \lambda_-}{2} + \mu_+ D^2, \quad
    p^{-+} = p^{+-} = \mu_- D^2, \\
    \langle \psi_{\pm}, \sigma \psi_{\pm} \rangle = 
    \lambda_{\pm} +  \mu_+.  \label{eq:15}
  \end{gather}
\end{theorem}

\par\noindent\textit{Proof.\ }
  We have to determine the commutant $H_{\psi_+}'$ of $H_{\psi_+}$. To this
  end note first that $H_{\psi_+,0} \subset H_{\psi_+}$ implies
  $H_{\psi_+}' \subset H_{\psi_+,0}'$. Hence consider the latter commutant
  first. By definition we have for each unitary $U$ on $\mathcal{H}_A \otimes  
  \mathcal{H}_B$
  \begin{equation}
    U \in H_{\psi_+,0} \Leftrightarrow [U,|\psi_+\rangle\langle\psi_+|]=0,\
    [U,\theta_A \otimes \idty]=0,\ [U,\idty\otimes \theta_B]=0,  
  \end{equation}
  where we have used the fact that the factorization $U=U_A\otimes U_B$ is a
  consequence of $[U,\psi_+]=0$; cf. \cite{VW1}. Therefore $H_{\psi_+,0}'$ is
  the von Neumann algebra generated by $|\psi_+\rangle\langle\psi_+|$,
  $\theta_A \otimes \idty$ and $\idty \otimes \theta_B$, i.e.
  \begin{equation}
    H_{\psi_+,0}' = \{ |\psi_+\rangle\langle\psi_+|, 
    \theta_A \otimes \idty, \idty \otimes \theta_B \}''. 
  \end{equation}
  By calculating all possible products of the generators this leads to
  \begin{multline}
    H_{\psi_+,0}' = \operatorname{span} \{ |\psi_+\rangle\langle\psi_+|,
    |\psi_-\rangle\langle\psi_-|, |\psi_+\rangle\langle\psi_-|,
    |\psi_-\rangle\langle\psi_+|,\\ P^{++}, P^{--}, P^{+-}, P^{-+} \}.
  \end{multline}
  The group $H_{\psi_+}$ is generated by $H_{\psi_+,0}$ and $V= V_A \otimes
  V_B$; cf. Proposition \ref{prop:1} Hence $A \in \mathcal{H}_{\psi_+,0}'$ is
  in $\mathcal{H}_{\psi_+}'$ iff it commutes $V$. Since $V_A$ and $V_B$ just
  exchanges the even with the odd subspace we easily conclude that
  \begin{equation}
    H_{\psi_+}' = \operatorname{span} \{ |\psi_+\rangle\langle\psi_+|,
    |\psi_-\rangle\langle\psi_-|, P^{++}+P^{--}, P^{+-}+P^{-+} \} 
  \end{equation}
  holds, which implies equation (\ref{eq:13}). Equations (\ref{eq:14}) and 
  (\ref{eq:15}) follow immediately from the definition of the $p^{jk}$ in
  (\ref{eq:9}) and from taking traces.
\hfill $\Box$ \medskip

If we decompose the $H_{\psi_+}$-invariant state $\sigma$ according to
Equation (\ref{eq:10}) we get
\begin{equation} \label{eq:18} 
  \sigma^{\pm\pm} = \frac{\lambda_+ +
    \lambda_-}{2 p^{\pm\pm}} |\varphi_{\pm}\rangle\langle \varphi_{\pm}| +
  \frac{\mu_+}{p^{\pm\pm}} P^{\pm\pm} \quad 
  \sigma^{\pm\mp} = \frac{P^{\pm\mp}}{D^2}.
\end{equation}
Hence if Alice and Bob perform $\theta_A, \theta_B$ measurements -- which they
can do without disturbing the systems -- they get either with probability
\begin{equation} \label{eq:32} p = p^{++} + p^{--}
\end{equation}
one of the (basically equivalent) isotropic states $\sigma^{++}$ or
$\sigma^{--}$, or they get with probability $1-p$ the totally chaotic state
$\sigma^{+-}$ or $\sigma^{-+}$. In case they get $\sigma^{\pm\pm}$ it is
distillable iff
\begin{equation} \label{eq:17} \langle \varphi_{\pm}, \sigma^{\pm\pm}
  \varphi_{\pm} \rangle > \frac{1}{D}
\end{equation}
holds. A straightforward calculation using Equations (\ref{eq:14}),
(\ref{eq:15}), (\ref{eq:18}) and (\ref{eq:17}) leads to the following
proposition 

\begin{theorem}{Proposition}
  Consider a state $H_{\psi_+}$ invariant state $\sigma$. The fidelity $f =
  \langle \varphi_{\pm}, \sigma^{\pm\pm} \varphi_{\pm}\rangle$ of the
  isotropic state $\sigma^{\pm\pm}$ from Equation (\ref{eq:18}) is given by
  \begin{equation} \label{eq:9} f = \frac{\langle \psi_+, \sigma \psi_+
      \rangle + \langle \psi_-, \sigma \psi_-\rangle}{p}.
  \end{equation}
  Hence $\sigma^{\pm\pm}$ is distillable iff
  \begin{equation} \label{eq:19} 
    \langle \psi_+, \sigma \psi_+ \rangle +
    \langle \psi_-, \sigma \psi_-\rangle > \frac{p}{D}
  \end{equation}
  holds.
\end{theorem}

Let us start with a general state $\rho$ and twirl over $H_{\psi_+}$, i.e.
\begin{equation}
  \sigma = \int_{H_{\psi_+}} U \rho U^* dU.
\end{equation}
Then $\sigma$ is $H_{\psi_+}$ invariant and Equation (\ref{eq:19}) is
equivalent to
\begin{equation} \label{eq:20} 
  \langle \psi_+, \rho \psi_+ \rangle + 
  \langle \psi_-, \rho \psi_-\rangle > \frac{p}{D}
\end{equation}
To get a general distillation protocol for Fermionic systems we can therefore
modify the distillation presented in Section \ref{sec:entangl-dist} as
follows:
\begin{enumerate}
\item 
  Start with $N$ \emph{distinguishable} copies of the same Fermionic
  system, each prepared in the same state (e.g. $N$ metallic wires containing
  an electron gas).
\item 
  Drop unentangled subsystems. This leads to $N$ bipartite Fermionic
  systems in the joint state $\rho^{\otimes N}$. Here $\rho$ is a density
  operator on $\mathcal{H}_A \otimes \mathcal{H}_B$, which should be
  interpreted, however, as a state of the algebra $\mathcal{A} \otimes
  \mathcal{B}$.
\item 
  Find maximally entangled states $\psi_{\pm}$ as in Equation
  (\ref{eq:16}) such that (\ref{eq:20}) holds.
\item 
  Average over the group $H_{\psi_+}$. This leads to the
  $H_{\psi_+}$-invariant state $\sigma$.
\item 
  Make $\theta_A, \theta_B$ measurement. If the outcome is $++$ or $--$ (which 
  happens with probability $p$)   this leads to the isotropic state
  $\sigma^{++}$ or $\sigma^{--}$. It can be treated with standard distillation
  techniques. 
\item 
  Otherwise ($+-$, $-+$) we get a chaotic state which is useless for
  distillation.
\end{enumerate}
To find a maximally entangled state $\psi_+$ such that (\ref{eq:20}) is
satisfied usually requires an optimization over all possible $\psi_+$. In
general this is very difficult. In the next section, however, we will discuss
a special class of states where this problem is more feasible and which
provide at the same time a systematic way of dropping unentangled modes.

\section{Quasifree states}
\label{sec:quasifree-states}

Let us apply the general scheme developed in the last section to quasifree
states. Recall that a density matrix $\rho$ describes a quasifree state of the
CAR algebra $\operatorname{CAR}(\mathcal{K})$ if there is a bounded operator
$S \in \mathcal{B}(\mathcal{K})$ such that
\begin{gather}
  \operatorname{tr}(\rho B(f_1) \cdots B(f_{2n+1})) = 0\\
  \operatorname{tr}(\rho B(f_1) \cdots B(f_{2n})) = 
  \sum \operatorname{sign}(p) \prod_{j=1}^{n} 
  \langle \Gamma f_{p(2j-1)} , S f_{p(2j)} \rangle, \label{eq:38}
\end{gather}
holds for all $n \in \mathbb{N}$ and $f_k \in \mathcal{K} \oplus
\mathcal{K}$. The sum in (\ref{eq:38}) is taken over all permutations $p$
satisfying
\begin{equation}
  p(1) < p(3) < ... < p(2n-1) , \quad p(2j-1) < p(2j) 
\end{equation}
and $\operatorname{sign} (p)$ is the signature of $p$. The \emph{covariance
  operator} $S$ is selfadjoint and satisfies
\begin{equation} \label{eq:21} 
  \Gamma S \Gamma = 1 - S,\quad 0 \leq S \leq \idty .
\end{equation}
We can express the right hand side of Equation (\ref{eq:38}) as the
\emph{Pfaffian} $\operatorname{Pf}(\tilde{S})$ of the antisymmetric matrix 
$\tilde{S}$ with matrix elements $\tilde{S}_{jk} = \langle f_j, S f_k\rangle$
$k > j$. Note also that this definition works as well in infinite dimensions,
although we will restrict our discussion in this 
chapter again to the case $\mathcal{K}_A = \mathcal{K}_B = \mathbb{C}^d$,
$\mathcal{K} = \mathcal{K}_A \oplus \mathcal{K}_B$. In $\mathcal{K}_{A/B}
\oplus \mathcal{K}_{A/B}$ we will use the bases ($k=1,\dots,2d$)
\begin{equation} \label{eq:23} 
  e_{A/B}^{(k)} =
  \begin{cases}
    e_{A/B,1}^{(k)} =  2^{-1/2} (|k\rangle +|k+d\rangle) & \text{if $k\leq d$} \\
    e_{A/B,2}^{(k-d)} = 2^{-1/2} i (|k-d\rangle -|k\rangle) & \text{if $k> d$}
  \end{cases}
\end{equation}
where $|k\rangle$, $k=1,\dots 2d$ denotes the canonical basis in
$\mathbb{C}^{2d}$. If the decomposition into Alice- and Bob-subsystems is not
important we can also use relabeled version
\begin{equation} \label{eq:31} 
  e^{(k)} =
  \begin{cases}
    e_A^{(k)} & \text{for $k \leq 2d$}\\
    e_B^{(2d-k)} & \text{for $k > 2d$}
  \end{cases}
\end{equation}
where $k$ ranges now from $1$ to $4d$. The advantage of this basis is its
$\Gamma$ invariance. We can therefore write
\begin{equation}
  S_{kj} = \langle e^{(k)}, S e^{(j)}\rangle = \langle \Gamma e^{(k)}, 
  S e^{(j)} \rangle, 
\end{equation}
with $k,j = 1, \dots, 4d$. In the following we will identify with slight abuse 
of notation the operator $S$ with the matrix $(S_{jk})_{j,k}$ and write
\begin{equation}
  S = \left( \label{eq:22}
    \begin{array}{cc}
      S_{AA} & S_{AB} \\
      S_{BA} & S_{BB}
    \end{array} \right) =
  \frac{1}{2} \left( 
    \begin{array}{cc}
      \idty_A +i X & i Y \\
      -i Y^T & \idty_B + i Z
    \end{array} \right).
\end{equation}
These expressions should be interpreted as block matrices with respect to the
Alice/Bob split, e.g.\ $S_{AB}$ contains all matrix elements of the form
$\langle e_{A}^{(j)}, S e_{B}^{(k)} \rangle$, etc. Using Equation
(\ref{eq:21}) it is easy to see that $X, Y, Z$ are \emph{real} $2d \times 2d$
matrices, and that $Y,Z$ are antisymmetric.

For quasifree states the expressions and constructions from the last two
sections can be given quite explicitly in terms of covariance matrices. The
following list summarizes the most important examples (cf. \cite{QFFermiDist}
for more details, in particular for proofs)

\begin{itemize}
\item The \emph{probability} $p$ to get equal parities during a joint 
  $\theta_A, \theta_B$ measurement (cf. Equation (\ref{eq:32}) is given by
  \begin{equation} \label{eq:10} p = \frac{1 + (-4)^d \operatorname{Pf}(S -
      \idty/2)}{2}.
  \end{equation}
\item A quasifree state with covariance matrix $P$ is \emph{maximally
    entangled}, iff $P$ has (in the basis from Equation (\ref{eq:23})) the
  form
  \begin{equation} \label{eq:24} 
    P = \frac{1}{2} \left(
      \begin{array}{cc}
        \idty_A & i R \\
        -i R^T & \idty_B
      \end{array} \right),
  \end{equation}
  with a real orthogonal matrix $R$. The quasifree state thus given can be
  represented by a state vector $\psi_P \in \mathcal{H}$ with
  \begin{equation}
    \psi_P = \frac{1}{\sqrt{2}} (\varphi_+ + \varphi_-)
  \end{equation}
  with maximally entangled vectors $\varphi_{\pm} \in \mathcal{H}_A^{\pm}
  \otimes \mathcal{H}_B^{\pm}$. In other words $\psi_P$ is always of the form
  $\psi_+$ assumed in Equation (\ref{eq:16}).
\item The \emph{fidelity} between a quasifree state $\rho_S$ and a maximally
  entangled quasifree state $\psi_P$ is given by
  \begin{equation} \label{eq:12} 
    \langle\psi_P, \rho \psi_P\rangle =
    \operatorname{Pf}(\idty - S - P).
  \end{equation}
\item The quasifree state $\rho_S$ can be transformed by a local Bogolubov
  transformation $u$ into a \emph{normal form} $\rho_{\tilde{S}}$ (which is
  again quasifree) such that the off diagonal blocks $-i \tilde{S}_{AB}$ and
  $i \tilde{S}_{BA}$ of the covariance matrix $\tilde{S} = u S u^*$ become
  diagonal with positive eigenvalues. To see this consider the singular value
  decomposition $Y = u_A \Sigma_Y u_B^*$ of $Y$ and choose $u = u_A \oplus
  u_B$.
\end{itemize}

The general distillation scheme described in the last section comprises the
search for a $\psi_+$ such that Equation (\ref{eq:20}) holds. Since we can
always choose $\psi_+ = \psi_P$ for some $P$ satisfying (\ref{eq:24}) a good
strategy is optimize the expression in (\ref{eq:12}) over all such $P$. The
following theorem treats an important special case (cf. \cite{QFFermiDist} for
a proof).

\begin{theorem}{Theorem} \label{thm:1} Consider a quasifree state $\rho_S$
  with covariance matrix $S$ from Equation (\ref{eq:22}). Assume that $X=0$ or
  $Z=0$ holds, and that $Y$ is diagonal with non-negative eigenvalues (the
  latter can be done without loss of generality). The maximal fidelity of
  $\rho_S$ with a quasifree, maximally entangled state $\psi_P$ arises if the
  basis projection $P$ is given by
  \begin{equation} \label{eq:26} 
    P = \frac{1}{2} \left(
      \begin{array}{cc}
        \idty_A & i \idty \\
        -i \idty & \idty_B
      \end{array} \right),
  \end{equation}
  and its value is
  \begin{equation} \label{eq:41} 
    \langle\psi_P, \rho \psi_P\rangle =
    \prod_{j=1}^{n} \left(\frac{1+\lambda_j}{2}\right)^{m_j}
  \end{equation}
  where $\lambda_j$, $j=1,\dots,n$ denote the eigenvalues values of $Z$ and
  $m_j$ the corresponding multiplicities.
\end{theorem}

If the condition $X=0$ or $Z=0$ is not satisfied the optimality statement is
in general not true. For states, however, which are already close to a
maximally entangled, quasi free state $X$ and $Z$ have to be at least small
(otherwise the condition $0 \leq S \leq \idty$ is not satisfied). Hence in
this case the choice $\psi_P$ with $P$ from (\ref{eq:26}) should be close to
the optimum (provided $Y$ is diagonalized). Therefore the following
specialization of the procedure from the last section should provide a
reasonably good scheme for distillation from quasi free, Fermionic states.

\begin{enumerate}
\item Consider $N$ bipartite Fermionic systems, each of which in the same
  quasi free state $\rho_S$.
\item Choose bases for $\mathcal{K}_A \oplus \mathcal{K}_A$ and $\mathcal{K}_B
  \oplus \mathcal{K}_B$ such that the block-offdiagonal part $Y$ of $S$
  becomes diagonal (and with real positive entries). As already pointed out
  above this can be done locally by Alice and Bob without any communication
  (if $S$ is known to them).
\item \label{item:1} Drop all modes except those belonging to the $n$ highest
  singular values of $Y$. The number $n$ must be chosen such that Equation
  (\ref{eq:20}) holds with $\psi_+ = \psi_P$ where the basis projection $P$ is
  (in basis which diagonalizes $Y$) of the form (\ref{eq:26}). The maximally
  entangled state $\psi_-$ is then quasi free as well, i.e. $\psi_-=\psi_Q$
  with basis projection
  \begin{equation}
    Q = \frac{1}{2} \left( 
      \begin{array}{cc}
        \idty_A & -i \idty \\
        i \idty & \idty_B
      \end{array} \right)
  \end{equation}
  (please check yourself). Hence the two fidelities $\langle \psi_{\pm}
  \rho_S, \psi_{\pm}\rangle$ in Equation (\ref{eq:20}) can be calculated with
  (\ref{eq:26}). If the probability $p$ is unknown Equation (\ref{eq:20})
  should be used with the conservative choice $p=1$.
\item Average (twirl) over the group $H_{\psi_+}$, make a $\theta_A, \theta_B$
  measurement and proceed as described in Section \ref{sec:dist-from-ferm}
\end{enumerate}

Let us demonstrate this scheme with free fermions (without spin) hopping on a
one-dimensional regular lattice $\mathbb{Z}$ (lets call it a ``wire''). They
can be described by the CAR algebra
$\operatorname{CAR}(\mathrm{l}^2(\mathbb{Z}))$ and the dynamics is given
formally\footnote{$H$ is not well defined as an element of
  $\operatorname{CAR}(\mathrm{l}^2(\mathbb{Z}))$, because the sum does not
  converge in norm. It gives rise, however, to a well define derivation and
  therefore the notion of ground state is well defined too.} by the
Hamiltonian
\begin{equation} 
  H = \sum_{j \in \mathbb{Z}} \left( c_j^*
    c_{j+1} + c_{j+1}^*c_j \right).
\end{equation}
It admits a unique quasifree ground state $\varphi_0$ with covariance operator
$S$ given by
\begin{equation} \label{eq:11} 
  S= \mathcal{F}^{-1} \left(
    \begin{array}{cc}
      E & 0 \\
      0 & 1-E
    \end{array} \right) \mathcal{F}
\end{equation}
where
\begin{equation}
  \mathrm{l}^2(\mathbb{Z}) \otimes \mathbb{C}^2 \ni F \mapsto 
  \mathcal{F}(F) \in \mathrm{L}^2(S^1) \otimes \mathbb{C}^2, 
  \quad   \mathcal{F}(F)(x) = \sum_{j=-\infty}^\infty e^{inx} F_n 
\end{equation}
is the Fourier transform and $E \in \mathcal{B}(\mathrm{L}^2(S^1))$ the
projection to the upper half-circle \cite{MR0295702,MR810491}.

\begin{figure}[h]
  \centering
  \includegraphics[scale=.7]{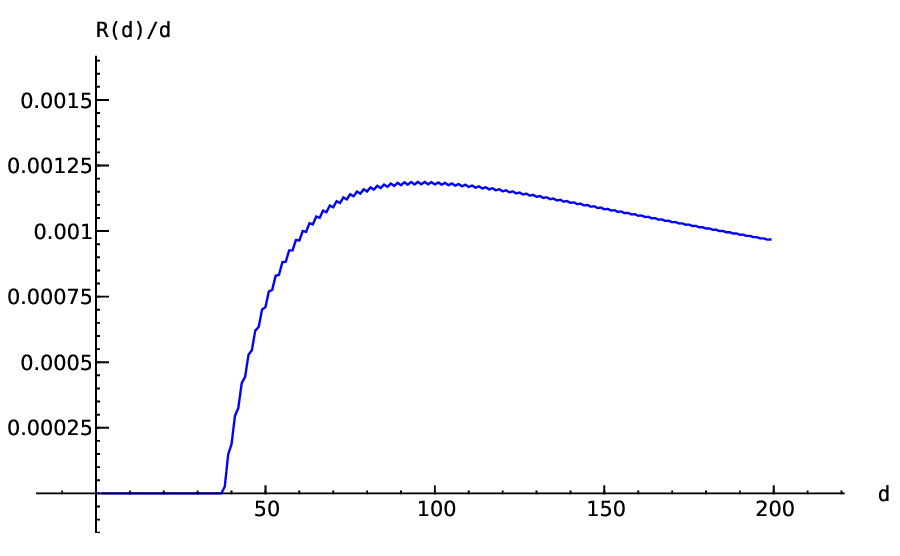}
  \caption{Distillation rate $R(d)/d$ for free Fermions on a one-dimensional
    lattice, if only two adjacent regions of length $d$ are accessed.}
  \label{fig:1}
\end{figure}

Now assume that Alice and Bob can control only two blocks of the form
$\Lambda_A = [0,d)$ and $\Lambda_B = [d,2d)$ (or any \emph{joint} spatial
translate of them). The restriction to the corresponding subsystem leads
exactly to the bipartite Fermionic system just studied. The reduced density
matrix $\rho_\Lambda$ arising from the ground state $\varphi_0$ is quasifree
and its covariance matrix in the basis (\ref{eq:31}) can be easily derived
from (\ref{eq:11}).

Now we can apply the distillation protocol given above. If we choose to keep
in step \ref{item:1} only the four highest singular values of $Y$ we get at
the end with probability $p$ from (\ref{eq:10}) a qubit pair in an isotropic
state $\sigma = f |\phi\rangle\langle\phi| + (1-f) (\idty -
|\phi\rangle\langle\phi|)/3$, with fidelity $f$ from Equation (\ref{eq:9}). If
a large number of systems is available (where system refers here to a whole
wire not to a single Fermion) and if the fidelity $f$ is big enough we can use
the Hashing protocol to distill maximally entangled qubit pairs. The
distillation rate, i.e. the number of maximally entangled pairs we get
asymptotically \emph{per wire} is \cite{VoWo03}
\begin{equation}
  R = p\bigl(1 - S(\sigma)\bigr) = p\bigl(1 + f\log_2(f) + (1-f)\log_2(1-f) -
  (1-f)\log_2(3)\bigr). 
\end{equation}
Maybe more interesting is the rate $R/d$ of pairs we get per \emph{lattice
  site} used. The result is plotted in Figure \ref{fig:1}. The small zigzag
noise on the graph arises from a slightly different behavior of the protocol
for even and odd values for $d$. This is an indication that the scheme is
indeed not optimal if the assumptions from Theorem \ref{thm:1} (i.e. $X=0$ or
$Z=0$) are not satisfied.

\end{document}